\documentclass[preprint,journal]{vgtc}

\ifpdf%        % if we use pdflatex
 \pdfoutput=1\relax     % create PDFs from pdfLaTeX
 \usepackage{graphicx}    % allow us to embed graphics files
 \DeclareGraphicsExtensions{.pdf,.png,.jpg,.jpeg,.PNG} % for pdflatex we expect .pdf, .png, or .jpg files
\else%         % else we use pure latex
 \usepackage{graphicx}    % allow us to embed graphics files
 \DeclareGraphicsExtensions{.eps}  % for pure latex we expect eps files
\fi%

\graphicspath{{figures/}{pictures/}{images/}{./}} % where to search for the images

\usepackage{microtype} 
\PassOptionsToPackage{warn}{textcomp} % to address font issues with \textrightarrow
\usepackage{textcomp}     % use better special symbols
\usepackage{mathptmx}     % use matching math font
\usepackage{amsmath}
\usepackage{float}
\usepackage{times}      % we use Times as the main font
\usepackage{cite}      % needed to automatically sort the references
\usepackage{tabu}      % only used for the table example
\usepackage{booktabs}     % only used for the table example
\usepackage[utf8]{inputenc}
\usepackage[english]{babel}
 
\usepackage[dvipsnames]{xcolor}

\onlineid{0}
\ieeedoi{10.1109/TVCG.2019.2934546}

\vgtccategory{Research}

\vgtcpapertype{application/design study}

\title{Cohort-based T-SSIM Visual Computing\\ for Radiation Therapy Prediction and Exploration}

\author{A. Wentzel, P. Hanula, T. Luciani, B. Elgohari, H. Elhalawani,\\ G. Canahuate, D. Vock, C.D. Fuller, G.E. Marai}
\authorfooter{
\item
 A. Wentzel, P. Hanula, T. Luciani, and G.E. Marai are at the University of Illinois at Chicago. E-mail: \{awentze2\,$|$\,gmarai\}@uic.edu
\item
 B Elgohari, H. Elhalawani, and C.D. Fuller are with the MD Anderson Cancer Center at the University of Texas. %\{BElgohari\,$|$\,HMElhalawani\,$|$\,CDFuller\}@mdanderson.org
\item
 G. Canahuate is at the University of Iowa. 
 \item
 D. Vock is with the School of Public Health at the University of Minnesota. 
}

\shortauthortitle{Wentzel \MakeLowercase{\textit{et al.}}: Cohort-based Visual Computing for Radiation Therapy}
\abstract{We describe a visual computing approach to radiation therapy (RT) planning, based on spatial similarity within a patient cohort. In radiotherapy for head and neck cancer treatment, dosage to organs at risk surrounding a tumor is a large cause of treatment toxicity. Along with the availability of patient repositories, this situation has lead to clinician interest in understanding and predicting RT outcomes based on previously treated similar patients. To enable this type of analysis, we introduce a novel topology-based spatial similarity measure, T-SSIM, and a predictive algorithm based on this similarity measure. We couple the algorithm with a visual steering interface that intertwines visual encodings for the spatial data and statistical results, including a novel parallel-marker encoding that is spatially aware. We report quantitative results on a cohort of 165 patients, as well as a qualitative evaluation with domain experts in radiation oncology, data management, biostatistics, and medical imaging, who are collaborating remotely. 
}


\teaser{
 \centering
 \includegraphics[width=\linewidth]{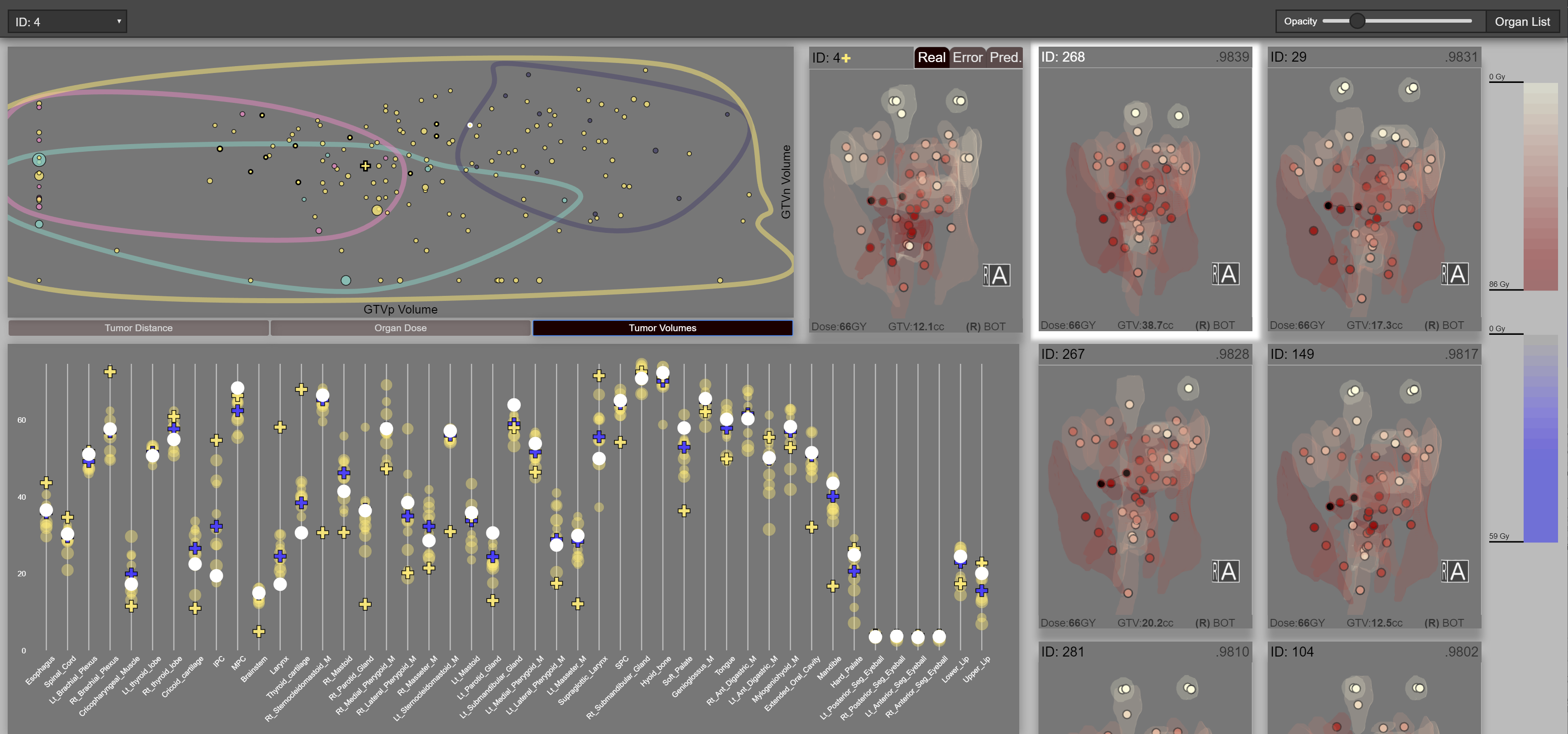}
 \caption{Visual computing for cohort-based radiation therapy (RT) prediction. A stylized 3D view of the predicted radiation plan of the current patient is placed centrally; top pale markers (front and back of eyes) receive the least radiation; tumors (black markers) receive the most. Additional RT views show the most similar patients under our novel T-SSIM measure, who contribute to the prediction; the most similar patient is currently highlighted (white) for comparison. A scatterplot (left) shows 4 clusters generated through the T-SSIM measure, with the current (cross) and comparison patient highlighted. A parallel-marker encoding (bottom) shows the predicted (blue cross) per-organ dose distribution within the context of the most similar patients; spatially collocated organs are in contiguous sections of the x-axis.}
 \label{fig:teaser}
}

\vgtcinsertpkg

\begin{document}

\firstsection{Introduction}
\maketitle

Modern radiation therapy (RT) has seen large advancements in the application of computational approaches for imaging and rendering structural data of a patient. However, once this information is extracted, the field requires a high level of human expertise and a tremendous amount of effort to create and develop personalized, high-quality treatment plans. For example, head and neck radiotherapy planning takes as long as a week, which, given that aggressive tumors double in 30 days, deteriorates the chances of tumor control and patient survival~\cite{Dan_Nguyen_et_al_2018}. Furthermore, radiotherapy plans also affect organs located nearby a tumor, resulting in significant toxicity (side effects) and loss of quality-of-life. There is no current method to predict toxicity before the development of the personalized plan.
%ensuing from a particular radiation plan based on the tumor location. 

With the emergence of large patient RT data repositories, there is growing interest in leveraging these repositories to computationally predict the dose distribution and toxicity for a patient before the actual RT plan is created. Under a healthcare model termed "precision medicine," such predictions would be based on outcomes registered for past patients with similar characteristics. These characteristics include the location of the tumor relative to the nearby organs at risk, which heavily influences the development of radiotherapy plans.

However, due to a lack of computational methodology to handle spatial similarity, radiation oncology clinicians rely solely on structural visual information from medical images, prior knowledge, and memory to guide the development of radiation plans and to forecast toxicity. This approach is not scalable.

In this work, we present a visual computing approach to RT planning, based on spatial similarity within a patient cohort. This approach introduces a novel spatial measure, T-SSIM, based on tumor-to-organs distance and organ volume, and its application in a novel predictive algorithm for dose distribution. The resulting algorithms are integrated with visual steering to support the algorithm development in a remote collaborative setting, as well as to derive insight into the role of spatial information. Specifically, the contributions of this paper are: 1) a novel hybrid topological-structural similarity measure for spatial data, inspired by an image fidelity technique; 2) the development of a predictive algorithm for RT dose distribution, based on this spatial similarity; 3) the design and implementation of an interface to guide the development of these algorithms, including a novel parallel-marker visual encoding which is spatially-aware; 4) the application of these algorithms and design to the emerging field of precision oncology RT planning, along with a description of this novel domain; 5) a quantitative and qualitative evaluation with collaborating domain experts.

\section{Related Work}
\label{sec:relwork}
Related work consists of other projects that deal with: spatial similarity measures; visual integration of spatial biomedical data with nonspatial data; and visual steering to assist in model development.

%%%MOVE TO INTRODUCTION
\noindent {\bf Spatial Similarity.}
%Our work is related to a large field in precision medicine, the practice of using information on a patient to improve treatment methods. 
Approaches in bioinformatics, pathology and oncology~\cite{Wenskovitch-2014-MAT,Gunduz,Petrakis,Kumar} facilitate spatial similarity by encoding spatial relationships through graph-based techniques. Unlike in our case, the underlying graphs are often small or constructed manually by clinicians~\cite{Zhang_P_et_al_2014,Fei_Wang_et_al_2011}. A second class of methods, based on 3D shape-based similarity, have been successful in shape retrieval applications in computer vision~\cite{iyer2005three, chen2003visual}. These methods typically experiment with artificial models such as CAD models or 3D scanner output, and focus on classifying models of very different shapes. These methods fall short of distinguishing anatomical objects within the same class, unless the objects have easily identifiable structures, such as the mandible and outer body contour~\cite{Chia-chi_Teng_et_al_2007, ruiz2005classifying}. In our case, structures are in the same class and do not have easily identifiable features. A third class of methods seeks to apply deep-learning to narrow versions of the similarity problem. For example, Nguyen et al.~\cite{Dan_Nguyen_et_al_2018} use deep-learning to predict dose distribution over a small set of organs in a cohort that had received the same type of RT plan, using tumor dosage and masks for organ 3D volumes. However, to date, no method has looked at automatically quantifying spatial similarity between patients for a large number of organs or a variety of treatments, or at presenting the prediction methodology in a way that can be understood by clinicians, as we do. % used the Hausdorff distance between anatomical structures for head and neck cancer patients to automatically identify lymph nodes.  While in our case templates do exist and are used to facilitate the identification of organs of interest in volume images, the large number of organs makes construction of  

\noindent {\bf Visualizing Biomedical Data and Nonspatial Data.}
Through established surface extraction and rendering algorithms, scientific visualization of biomedical data has been able to gradually shift its research focus towards visual computing~\cite{lorensen2004death}, integration of nonspatial data~\cite{C._Johnson2004}, and new technologies. For example, instead of rendering magnetic resonance data from scratch, Nunes et al.~\cite{nunes2014an} focused on analysis, by linking existing medical imaging software (MITK~\cite{wolf2004medical}) with statistical views of metabolic data to support delineation of target volumes in RT planning. In recent RT plan visualization research, Patel et al.~\cite{Daniel_Patel_et_al_2007} use virtual reality (VR) to visualize RT plans, allowing 3D structure visualization with hue and opacity, as typically done in desktop applications. Ward et al.~\cite{Ward_JW_et_al_2007} describe a VR system for radiation planning that allows the user to alter beam positions. Although these and other works have led to advances in viewing and planning specific radiotherapy plans in detail, none of these works seek to compare RT plans between patients or make predictions. Two other works~\cite{raidou2016visual, raidou2018bladder} have proposed visual tools for the exploration of uncertain tumor control probabilities in the prostate, and dose delivery accuracy as a function of bladder shape analysis, respectively. These works do not consider spatial similarity, surrounding organs at risk, or the RT plan as a whole. %Additionally, these methods are computationally expensive, which creates issues for remote collaboration and accessibility when using the tool.

In terms of spatial-nonspatial data integration, two prevailing paradigms for integrating spatial and non-spatial features exist: overlays and multiple coordinated views (sometimes called linked views)~\cite{Marai-2015-VSI}. In biomedical scientific visualization, an overlay approach~\cite{ten2007functional,ten2008data,bottger2014three} is commonly used when the non-spatial feature is scalar. As the non-spatial data becomes more complex (connectivity, clusters, dynamic characteristics, other statistics), the linked-view paradigm~\cite{jianu2009exploring,beyer2013connectomeexplorer,al2016neuroblocks} becomes prevalent. Several reports~\cite{maries2013grace,Marai-2015-VSI,G._Elisabeta_Marai_et_al_2019} further support the use of coordinated views in collaborative tasks which involve multiple users with complementary expertise.
Other more recent approaches~\cite{nowke2013visnest,maries2013grace,ma2018rembrain} use a hybrid approach that consists of both overlays and linked views. We follow a similar hybrid approach to support the exploration of RT plan data.

\noindent {\bf Visual Steering for Model Development.} 
Visual steering (or integrated problem-solving environments) is a top problem in scientific visualization~\cite{C._Johnson2004}. Under this research umbrella, visualization tools for predictive model development have been developed for domain-specific applications. Naqa et al.~\cite{I_El_Naqa_et_al_2006} built a visualization tool to help create statistical models for dose-toxicity outcomes for specific organs, using a combination of statistical views and model controls. Unlike our work, their project assumed that the dose-distribution was already known, and was restricted to individual organs. Poco et al.~\cite{Jorge_Poco_et_al_2014} built a system for visualizing and developing similarity measures in environmental data, but focused on abstracted views for improving the measures without referring to underlying spatial patterns, as we do. Kwon et al.~\cite{Bum_Chul_Kwon_et_al_2018} provided a generic method for clustering model development, and used it for the development of patient similarity when diagnosing heart failure, but with no spatial data included. Visual steering tools based on multiple coordinated views appear also in visual encoding design~\cite{Kresimir_Matkovic_et_al_2008}, engineering~\cite{J_Waser_et_al_2010, Ribii_H_et_al_2013}, epidemiology~\cite{maries2013grace}, cell signaling~\cite{smith2012rulebender}, and artificial intelligence~\cite{Sean_McGregor_et_al_2015}; some of these works emphasize visually adjusting a simulation as it progresses, while others couple the steering with off-line processes. These methods differ from our goals in the key consideration of the problem space. We are interested in developing predictive models using RT medical data, which has unique requirements related to spatial and statistical data.

\section{Methods}
\subsection{Domain Background and Problem}
In head and neck cancer treatment, RT is often used as a primary or secondary treatment for patients. Radiation oncology relies heavily on the use of imaging in order to obtain information about the patient's tumor and surrounding organs. Traditionally, data acquisition is accomplished via magnetic resonance imaging (MRI), computational tomography (CT), or ultrasound. These techniques provide 2D image slices across the target volume, that can then be segmented to identify organs of interest, and used in diagnostics and radiotherapy planning. Current planning techniques typically use these images overlaid with a color map, allowing clinicians to 'paint' the dose across the organ as a way to visualize the outcome of the different radiation plans ~\cite{Uulke_A._van_der_Heide_et_al_2012}.

In radiation therapy planning, a primary concern is limiting dose to organs at risk near the target volume, while maximizing tumor exposure. For example, a head and neck tumor may receive 66-72 Gy units of radiation, while nearby organs at risk ideally would receive lower amounts. Unfortunately, that is not always possible, and radiotherapy has been linked, by several studies, to organ damage and long-term toxicity (side effects), including xerostomia (permanent dry mouth), and swallowing complications ~\cite{C_LING_et_al_1996, Maria_Werner-Wasik_et_al_2010, Hale_B._Caglar_et_al_2008}. In light of these considerations, high-precision methods have been developed that allow for complex, highly conformable radiotherapy plans to be developed and delivered. Intensity-modulated radiation therapy (IMRT) is one such method.

IMRT allows for delivering more precise dose distributions via multiple (5-9) different radiation beams, each with tunable intensity distribution\cite{C_LING_et_al_1996}. The increased complexity of these plans comes at the cost of longer planning time and heavy reliance on clinician knowledge~\cite{William_M._Mendenhall_et_al_2006}. IMRT plans are typically created through a mixture of expert knowledge and planning software, with repeated trial-and-error rounds and consultation between the planner expert and the physician regarding the plan quality and tradeoffs. Beams are typically set up at an expert-determined fixed height, in order to reduce the problem space for the optimization software. Constraints are given on the allowable doses to organs of interest, which typically take the form of maximum doses to organs at risk, and minimum constraints to the target dose~\cite{S_Webb1994}. As a result, the problem space and planning time are expensive, and there is keen interest in leveraging computational techniques to support predicting the outcome of the radiation plan earlier in the process. 

At the same time, the high incidence of cancer cases has led to the creation of large repositories of patient data, along with their diagnosis scans, their respective RT plans, and treatment outcomes. Under the "precision medicine" healthcare model, practitioners seek to leverage these repositories in order to predict, for a specific patient, the most appropriate therapy course, along with the outcomes of that treatment. Unlike in personalized medicine, the precision medicine prediction is based on data collected from a cohort of similar patients in the repositories~\cite{G._Elisabeta_Marai_et_al_2019}. 

While cohort similarity based on abstract data (e.g. genetic sequence profile) is in general well researched in the statistics community, there is a general lack of spatial similarity methodology. In the domain characterization discussed in this work, our radiation oncology collaborators would like to be able to automatically retrieve, given the diagnostic scan of a new patient, a cohort of patients with similar tumor location. Currently, this is done based on clinician or institution memory alone, which is not scalable. Should such an automated similarity measure become available, the domain experts would then like to analyze the patterns in the RT plans of the patients within that cohort. Based on that information, they would like to predict the RT dose distribution for the new patient and its potential effects, without going through a detailed RT planning process from scratch. Because these tasks and activities rely on the visual assessment of spatial similarity and prediction in terms of dose distribution over the head and neck organs, the problem stands to benefit from a visual computing solution.

We arrived at this domain characterization of precision RT planning through a two-year collaboration with a team of radiation oncologists and statisticians located at multiple geographical sites. During this collaboration, we (four visual computing researchers) held weekly remote meetings and quarterly in-person meetings with a group of four radiation oncologists, a data management specialist, and a statistician. To characterize this novel application domain and design a solution, we followed an Activity-Centered-Design paradigm (ACD) as described by Marai~\cite{8017610}, coupled with team science principles for remote collaboration, previously described~\cite{G._Elisabeta_Marai_et_al_2019}. 

\subsubsection{Design Process}
We implemented the theoretical ACD paradigm through an iterative, multi-stage process. After identifying and confirming with our collaborators the main activities to be performed, the research team met weekly with the domain experts, as the algorithms and application were being developed and the design refined, to collect feedback, and to verify that evolving requirements were being satisfied.  In concordance with ACD, we used a quantitative methodology to assess the capabilities of the resulting solution, and a qualitative evaluation methodology with note-taking to analyze the user activities.

\subsubsection{Data processing}
The cohort data for this project is part of a repository of head and neck cancer patients from the MD Anderson Cancer Center that have received IMRT. Contrast-enhanced computed tomography (CECT) volume imaging data from the initial patient diagnoses were retrieved through commercially available contouring software~\cite{Varian_Medical_Systems_2018}. Contours were manually segmented to extract primary (GTVp) and secondary nodal gross tumor volumes (GTVn), as well as volumes of interest in the prediction related to organs at risk. Each CECT image was 512$\times$512 pixels, with a slice thickness of 1.25-5mm. Connected tumor volumes were treated as one volume. After segmentation, we used a custom Matlab script to extract a list of structural features for each volume of interest: volume, centroid position, and distance between each volume of interest, including tumors. Distance was measured as the minimum distance between the two volumes. Dosimetric data on the minimum, mean, and maximum dose for each volume of interest was extracted from radiation plans. Additional data on each patient's treatment plan was also included, which included the patient's tumor laterality, tumor subsite, and prescribed dose. All patient data was anonymized; patients were coded using dummy IDs.

45 organs of interest were identified as being of interest by our oncology collaborators, in addition to the primary and secondary tumor volumes. Of the candidate patients, only those with data on all 45 organs, and at least one primary or secondary tumor volume were included. Since segmentation and labeling of the data were done manually for higher accuracy, some anomalies in the dataset were found after visual analysis. Patients with organ position or mean doses more than 3 standard deviations about the population average were flagged and analyzed alongside our collaborators using the visual computing solution, and those with likely erroneous radiation plans were also excluded. The selection criteria was demographics-agnostic to prevent selection bias. 165 patients (140 male, age 59+-8.75 years, tumor N-staging~\cite{amin2016ajcc} 0th through IVth distribution: 32, 18, 91, 6, and 18, respectively) out of 245 candidate patients were included in the final cohort. The data could be further filtered based on patient characteristics. Data was then post-processed in order to compute derived features used in the visual interface, create dose predictions, and label patients with clustering results, as described below.
%This process included identifying the main expert activities to be performed as part of the visual analysis process. The activity-design process further included fact-based semi-structured interviews with the domain experts, to confirm functional requirements related to the activities to be supported by the system, and to explore prototypes. No data about the domain expert individuals was collected. Once we identified the main problem and the data used in this project, and further following the ACD theoretical model, we confirmed the result of our analysis with our medical collaborators.
\begin{figure*}[th]
 \centering 
 \includegraphics[width=\textwidth]{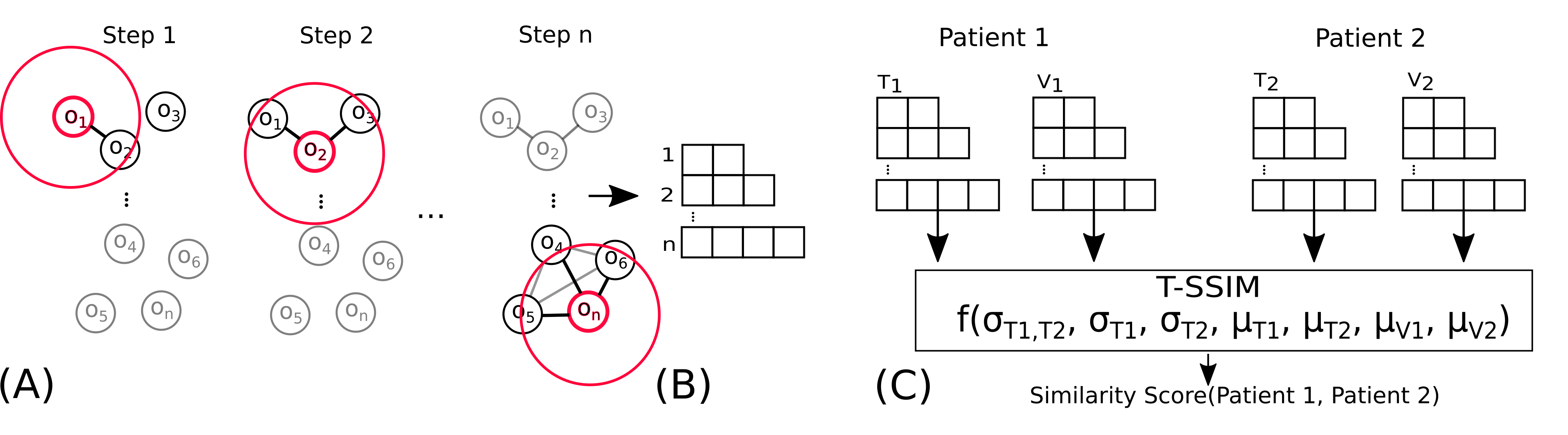}
 \caption{Construction of the spatial similarity measure. (A) A sliding window (a sphere, illustrated in 2D here) steps through the centroids of the organs, to identify nearby organs. (B) Each step in the sliding window constructs a variable-length vector based on the set of nearby organs (e.g., 2 organs in Step 1, 3 in Step 2, 4 in the n step). (C) We create two sets of vectors populated with tumor-organ distances and volumes, respectively, for each patient.  These vectors are used as inputs into a similarity function (T-SSIM) to compare two patients. The vectors can be represented in matrix form (Subsection~\ref{sec:tssim}).}
 \label{fig:algorithm}
\end{figure*}

\subsection{Algorithms}
In order to support computing over images and 3D models (i.e., visual computing) for this project, we need to design appropriate algorithms for spatial similarity and prediction, described below.

\subsubsection{T-SSIM Spatial Similarity Algorithm}
\label{sec:tssim}
In constructing a similarity algorithm special considerations need to be made for our problem. First, traditional methods of measuring similarity along feature vector representations, such as correlation or mean-squared-error, do not take into account the original structure inherent in the patient's anatomy. Second, neither shape-based techniques nor deep-learning techniques are a good match for this problem (Section~\ref{sec:relwork}). Third, the large number of organs-at-risk considered, and the lack of clinician agreement make infeasible the manual construction of a 3D graph structure based on the head and neck data. Fourth, an algorithmically constructed 3D graph-structure would have large edge cardinality, making graph-based matching algorithms infeasible. Because of these considerations, we arrived at a hybrid solution: 1) construct a topological structure based on organ adjacency; this structure will be common among all patients; 2) for each patient, generate two copies of the structure with tumor-to-organ distance data and volume data, respectively, specific to that patient; 3) define a similarity measure over these patient-specific data structures, inspired by image processing. Figure~\ref{fig:algorithm} illustrates this process.

%Conversely, automated cross-registration of whole head and neck volume images is not reliable; most organs at risk are manually segmented to ensure adequate accuracy. Several other standard approaches to incorporating spatiality did not lead in our case to meaningful results: several attempts at defining and using 3D graph similarity based on anatomical location yielded high similarity scores across the cohort, which was not meaningful. Further attempts at directly clustering the organs based on centroid location and dimensionality reduction lead as well to high similarity, meaningless scores across the board. This makes sense, given that most humans have similarly positioned organs, which biases the match.
Our spatial similarity algorithm is inspired by the Structural Similarity Index (SSIM)~\cite{Z._Wang_et_al_2004}, which is traditionally used to measure signal fidelity when comparing two images. Since the SSIM was designed for image processing, it takes advantage of an important assumption about the data: that pixel position serves as a direct analogue of spatial position. Because our data is already a reduced set of features (organs and tumors), rather than the original CECT images, this image-based assumption no longer holds. However, by reformulating the problem, we can use the spatial data we have to achieve the same effect, as described below. We refer to this novel reformulation as the Topological Structural Similarity Index, or T-SSIM.

In the original SSIM, a sliding window is used to calculate image similarity between the same local regions in two images. This local similarity is computed as:
$$SSIM(A,B) = \frac{(2 \mu (A) \mu (B) + c_{1})( 2 \sigma (A,B) + c_2)}{ (\mu (A) ^{2} + \mu (B)^{2} + c_{1})( \sigma (A,A)^{2} + \sigma (B,B)^{2} + c_{2})}$$
%\label{equ:one}
\noindent where $\mu (A)$ is the mean of matrix A, $\sigma (A,B)$ is the matrix covariance between two matrices A and B, and $\sigma (A,A)$ is the self-covariance of matrix A; $c_{1}$ and $c_{2}$ are small constants that are used for numerical stability. One of the reasons SSIM uses a local window is because image features and distortions are often space-variant. The window serves to isolate pixels within a certain distance from each other, so window size serves as a direct analog for actual distance. In contrast, our data is spatially bound to the centroids of each target volume. Thus, we need to find a way to encode the distance between the centroids, rather than a pixel distance. While the direct equivalent of a sliding window would be constructing a 3D area and sliding through different voxels, most of those voxels would be empty. Instead, we construct a topological equivalent.

In order to construct a topological equivalent to the SSIM image data, and create a sliding window analog, we need notation to describe when two volumes are within a window, for which we will use the concept of spatial adjacency. Let us define a matrix $\overline{D}^{\mid O \mid \times \mid O \mid}$, where $d_{i,j} \, \in \, \overline{D}$ denotes the average distance between organs $i$ and $j$ across the cohort. We define two organs as being adjacent when the average distance between them is less than a certain distance $d_{max}$. Mathematically, we can write this as $o_{j}\, \sim \, o_{i} \, \forall \, o_{j}, \, o_{i} \, \in \, O \, \mid \, d_{i,j} < d_{max}$, where the $\sim$ operator denotes adjacency. If we consider our window to be a 3D sphere centered at a point, we can define all organs within the window as all the points adjacent to the center of the sphere (Fig.~\ref{fig:algorithm}A). For efficiency, we will only consider the set of windows centered at each organ. Conversely, we can represent this set of windows as an adjacency matrix $M^{\mid O \mid \times \mid O \mid}$:
\[M_{i,j} = 
 \begin{bmatrix}
 1 & o_{i} \sim o_{j}\\ 
 0 & else
 \end{bmatrix}
\]
In other words, the row $M_{i}$ in our topological structure is a row representing all organs that are within a certain distance from organ $i$ (Fig.~\ref{fig:algorithm}B). Via line search~\cite{swann1969survey} so that the whole topological structure is connected, we found the optimal parameter $d_{max}$ as 80mm for the window size. The topological structure is common across all patients.

The next element we need is a pixel value analog. In our data, each organ is bound to several variables that could be used. Alternatively, we can compute similarity over multiple variables, and take a weighted average of them. The downside of such an approach would be that not all possible variables influence equally the final result, so using multiple values would require careful weighing of the values. To overcome this problem, we consider the underlying formulation of the SSIM. 

The original SSIM formulation can alternatively be written as the composition of three functions for intensity (luminance), contrast, and structure. These components can be written as:
$$l(x,y) = \frac{2 \mu(x)\mu(y) + c_{1}}{\mu(x)^{2} + \mu(y)^{2} + c_{1}}$$
$$c(x,y) = \frac{2 \sigma(x)\sigma(y) + c_{2}}{ \sigma(x)^{2} + \sigma(y)^{2}}$$
$$s(x,y) = \frac{2 \sigma(x,y) + c_{2}}{2 \sigma(x) \sigma(y) + c_{2}}$$

\noindent using the same SSIM notations. This formulation allows us to combine multiple variables. While we found that the distances between the primary tumor and each organ provided good matches using the original SSIM formulation, we can augment that measure by considering the organ volume as another intensity channel. 

For notation, let us consider the set of the organs adjacent to organ $i$, $M_{i}$, and patients A and B. Let us instantiate a copy of the topological structure with the matrix of tumor-organ distances $T^{\mid P \mid \times \mid O \mid} $ and another copy with the matrix of organ volumes $V^{\mid P \mid \times \mid O \mid}$ (Fig.~\ref{fig:algorithm}C), where $T_{i,j}$ represents the $j$th organ of the $i$th patient. We want to perform calculation over subsets of adjacent organs that we encoded in $M$. We can write each of these local subsets of values as $M_{i} \cdot T_{j} = T_{j}^{(i)}$ and $M_{i} \cdot T=V_{j} = V_{j}^{(i)}$. Put simply, $T_{j}^{(i)}$ is the set of tumor-organ distances for all the organs near organ $i$, for patient $j$. With this notation, we can now define local similarity as:

$$f_{i}(A,B) = l(T_{A}^{i}, T_{B}^{i}) \, l(V_{A}^{i}, V_{B}^{i}) \, c(T_{A}^{i}, T_{B}^{i}) \, s(T_{A}^{i}, T_{B}^{i})$$

By summing up the local similarity scores along the entire set of organs, we obtain a similarity score for patient A and patient B.  We can then generate a matrix of similarity scores $S^{\mid P \mid \times \mid P \mid}$, where each entry is:

$$S_{A,B} = \frac{\sum_{i=0}^{\mid O \mid} f_{i}(A,B)}{\mid O \mid}$$

Scores are normalized across the dataset to be between 0 and 1. In Fig.~\ref{fig:teaser} right, note how this measure successfully retrieves patients with similar tumor location.

\subsubsection{Prediction and Statistical Analysis}
To predict a patient's dose distribution, we use a weighted k-nearest-neighbors algorithm, which is a common method of prediction in similarity-based health models~\cite{Anis_Sharafoddini_et_al_2017}. The dose distribution prediction was calculated as the per-organ dose average of the k most similar patients:

$$Rad_{i,j}^{predicted} = \frac{\sum_{n \, \in \, N_{i}} S_{n,j}Rad_{n,j}}{\sum_{n \, \in \, N_{i}}S_{n,j}}$$

\noindent where $Rad^{\mid P \mid \times \mid O \mid}$ is a matrix of radiation doses across the cohort, $Rad_{i,j}$ denotes the radiation dose to the $j$th organ for the $i$th patient, and $N_{i}$ is the set of the k most similar patients to patient $i$.

%The resulting square matrix $Rad$ has dimensions equal to the number of patients in the cohort.

Even before applying clustering to this similarity matrix, we started noticing unusual groups of patients forming based on this similarity measure, and specific patterns of radiation distribution. An immediate goal became to perform clustering and statistical analysis using this spatial measure, and incorporate the resulting information: each patient was labeled with a cluster computed separately from the similarity measure, as discussed in Section~\ref{sec:evaluation}. To allow for the dosimetric and tumor-organ distance data to be viewed across the whole dataset, principal component analysis (PCA)~\cite{_Principal_Component_Analysis_and_Factor_Analysis} was done on the matrix of radiation doses $Rad$ and tumor-organ distance $T$. When making a prediction, only patients within the same cluster were considered. When analyzing the optimal number of matches for our prediction (Section~\ref{sec:evaluation}), we found that the number varied with the size of the cluster, and making the parameter tunable for different clusters helped improve performance. After testing different parameters via line search~\cite{swann1969survey}, we found that a good number of matches to use was the square-root of the cluster size.

Because the input RT plans already consider maximum organ doses, and minimum target constraints~\cite{S_Webb1994}, the predicted results fall within clinically acceptable ranges. All data processing, calculations for similarity, predicted dose, and PCA were computed offline, and information was exported as a JSON file for use in visual steering.

\begin{figure*}[t]
 \centering 
 \includegraphics[width=0.8\textwidth]{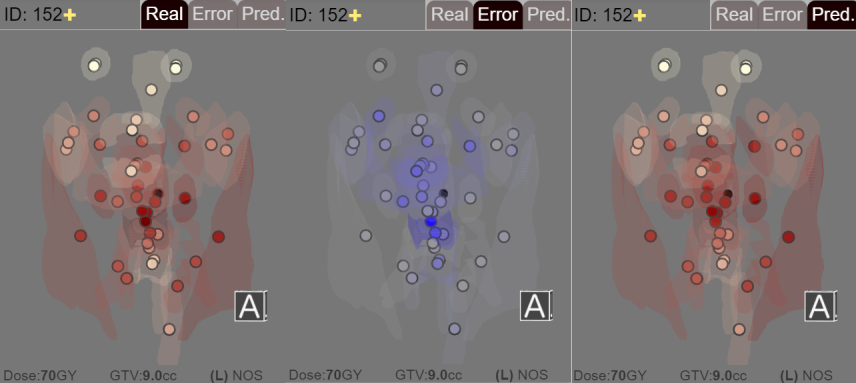}
 \caption{Three stylized views of the 3D radiation plan for Patient 152 showing the actual (left), the predicted (center), and the prediction error (right, in blue) in the radiation plan. Circular markers indicate the location of organs at risk, and black markers indicate the tumors. Red luminance is mapped to the radiation dose (higher dose mapped to darker shades) and blue luminance is mapped to error size. Transparent organ models are shown for context. The pale markers at the top correspond to the eyes, and the lowest marker is located down the spine.}
 \label{fig:rtplans}
\end{figure*}

%%%\begin{figure}[h]
%%% \centering 
%%% \includegraphics[width=\columnwidth]{matches}
%%% \caption{Scrollable panel showing the radiation plans of 4 patients with the highest similarity score to a selected patient. A scale on the right shows a visual reference of the mapping between dose and color used to color the radiation plans. }
%%% \label{fig:rtmatches}
%%% \end{figure}

\subsection{Visual Steering Design}
Once the visual computing algorithms are defined, a visual analysis interface enables the domain experts to steer the further development of these computation processes. By introducing an interactive visual steering component, we are able to leverage domain-specific knowledge, and support the discovery of patterns in the data.  The visual analysis component of this application (Fig.~\ref{fig:teaser}) followed multiple design iterations, aligned with the similarity algorithm and prediction algorithm development. The final prototype design was designed to support the following activities (i.e., sets of tasks), derived from the domain characterization: (1) analyze the result of data clustering and similarity measures in the context of the entire cohort, and of spatial and dosimetric data, (2) analyze the inherently spatial dosimetric data extracted from the patients' scans and radiation therapy plans in a way that is visually intuitive to the domain experts, (3) compare those similar patients used in dose predictions, (4) analyze the result of our T-SSIM patient similarity measure, and (5) analyze the results of the dose prediction algorithm.

The final prototype comprises several coordinated views. We chose to use linked views because they allow visual scaffolding from familiar visual representations to less familiar encodings~\cite{Marai-2015-VSI}. Unlike public health research, which is focused on cohorts, precision medicine is about the treatment of a specific patient, so the entry point to the application is a search box for a specific patient within the cohort (the default is the first patient). Because radiation oncologists are familiar with RT plan renderings, a 3D stylized radiation plan of the selected patient is placed centrally on the screen (activities 2 and 3). Additional RT views for the most similar patients give the patient a local context, and allow users to assess how the prediction algorithm is being used concretely (activities 3, 4, and 5). To support analysis within the cohort, and allow for clustering studies context (activity 1), a scatterplot shows the clustering data among different dimensions that can be explored. Finally, we provide a novel encoding that allows for the local dose distribution of each organ of interest to be understood within the context of the k most similar patients (activities 3, 4, and 5). By linking the views, we provide a way of allowing specific plans to be understood within context, and we support a variety of workflows for exploring the data. We describe each component in detail below.
 
\subsubsection{Stylized Radiation Plan Renderings}
Centrally in the visual interface is a stylized 3D rendering of the radiation plan for the selected patient (Fig.~\ref{fig:rtplans}). Organs of interest are represented as circles drawn at the organs' centroids. In order to reduce issues with segmentation and allow the visualization to be rendered without requiring information on the entire 3D contours from the CECT scans, the organ shapes are represented using transparent, generic 3D VTK models, centered at the centroids of the target volumes. A slider changes the opacity level of the organ models in the radiation plan, as well as the color-scale to the right of the radiation plans. By combining centroid data and generalized models instead of full 3D contours, we effectively reduce the computational requirements of the system and minimize visual occlusion while still showing a recognizable 3D structure of the patient anatomy. We encode dose to each organ with the luminance of the respective centroid node and model; we encode larger doses with darker values. Gross Tumor Volumes (GTVs) are shown only as nodes located at the tumor centroids, drawn in black, to make then identifiable, as there are no corresponding 3D contours for these regions. Additionally, when both a primary tumor (GTVp) and secondary (GTVn) are present, a line segment is drawn between these nodes, to further emphasize their spatial relationship. These stylized 3D views, as well as a miniature cube with orientation labels (scene bottom-right corner), can be rotated in sync by direct manipulation to allow the user to more easily see specific areas while still being able to quickly recall the current orientation. Additional marks, labels, and details on demand display information about organ names, dosage, volume, and tumor location, to help correlate information across the views. This stylized 3D view was the result of several design iterations, ranging from highly stylized node-link renderings of the organs to fully-fledged volume renderings, and a variety of markers and labels to indicate current orientation and details.

Because one of the goals is to be able to analyze the result of the prediction algorithm, tabs above the radiation plan allow the user to change the view to the predicted plan, and to the prediction error in the plan. We encode the prediction error using a blue hue in order to distinguish which information is currently shown.

A separate, scrollable panel (Fig.~\ref{fig:teaser} right) shows similar stylized 3D views for the nearest-neighbors of the selected patient, sorted by descending similarity. Allowing the user to control the matched radiation plans separately supports the placement of those plans near the selected patient plan, for easier comparison. For these neighbor RT plans, the similarity score between the given patient and the currently selected patient is shown in the top-right corner. Two color scales, automatically populated to encode the upper bounds of the doses found in the dataset, serve as a visual reference for colormaps, as well as inform the user of the minimum and maximum mean dose, and prediction error in the data. A neutral 18\% gray background was used to allow for better contrast with the transparent colored visual encodings used in the RT views, at both high and low values~\cite{giesen2007conjoint, schloss2018mapping, bartram2010whisper}. %A neutral gray background was used to allow for contrast with both colored visual encodings and black text~\cite{evergreen2013design}, and to allow for white to be used for brushing and linking.

\begin{figure}[ht]
 \centering 
 \includegraphics[width=\columnwidth, height = 90mm]{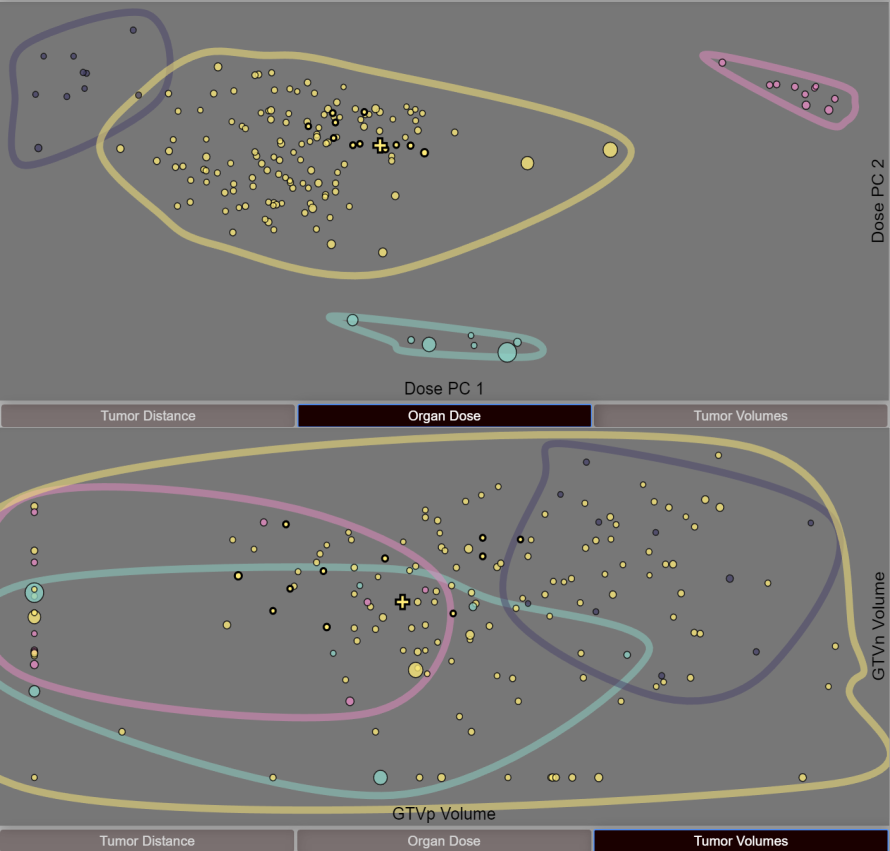}
 \caption{Two configurations of the scatterplot. The data can be plotted across the principal components of the radiation doses (top), primary and secondary tumor volumes (bottom), and principal components of the distances between each organ and the primary tumor volume (see Fig.~\ref{fig:teaser} top left).}
 \label{fig:scatterplots}
\end{figure}

\subsubsection{Scatterplot View}
A main activity of interest to our collaborators was being able to analyze clustering results in the data. Additionally, we wanted a way to find correlations across the dataset, to help identify where the largest prediction errors were occurring. Since the main data of interest was the relationship between spatial information and the radiation plan, followed by dose prediction, we selected the distances between the GTV and the 45 organs of interest and the dose information, respectively, as two of the feature spaces that could be viewed. For these feature spaces, PCA was done to project the 45 data dimensions to two. After several visual computing iterations and further discussion with collaborators (described in the Evaluation section), it was determined that tumor volume was also an important factor, and so it was included as an additional space. Since tumor volumes are usually categorized in 3-4 discrete stages, we used both the GTVp and GTVn volumes as proxy values to allow for better discrimination among the cohort. 

Patients in the scatterplot are color-coded according to cluster labels. The number of clusters shown was decided also through several design iterations, described in Section~\ref{sec:evaluation}. In order to allow for easier perception of outliers, an envelope is drawn around each cluster. Animated transitions when changing the axis variables in the scatterplot allow for a visual understanding of how the different clusters are distributed across multiple dimensions (Fig.~\ref{fig:scatterplots}). Tooltips on the scatterplot allow the user to view the name, size, mean dose, and mean prediction error for the entire cluster. 
 
Markers in the scatterplot are sized by the error in the radiation dose prediction for each patient to allow easy identification of patterns in prediction error, and to find outliers in the data. By default, patients are represented as semi-transparent circular markers, while a different shape is used for the patient in focus (a cross) so they can be more easily identified via pre-attentive cues. In an application of Tufte's layering and separation principle~\cite{Tufte:1990:EI}, patients used as matches for the selected patient are given a higher opacity and larger border so that they can be identified among the rest of the cohort. Additional tooltips allow the user to view the patient ID, position, mean dose, prediction error, cluster, and current position in the scatterplot.

%%%\begin{figure*}[h]
%%% \centering 
%%% \includegraphics[width=\textwidth]{DoseErrorPlot}
%%% \caption{Parallel mark plot of the dose distribution the for each organ in the current selection. Dose prediction error can be judged by looking at the distance between that organ's cross marker (actual dose), and blue cross marker (predicted dose) for each organ. Spatially collocated organs are in contiguous sections of the x-axis.}
%%% \label{fig:doseexplorer}
%%%\end{figure*}

%%%\begin{figure}[h]
%%% \centering 
%%% \includegraphics[width=\columnwidth]{linkingSmall}
%%% \caption{Composite shot showing view linking via color, markers, and brushing. Linked encodings include the markers in the scatterplot (top left), the ID and marker above a patient's 3D radiation plan (Fig.~\ref{fig:teaser}), the selected match in the RT similarity panel (top right) and markers in the radiation dose plot (bottom). Hovering the mouse over any of these encodings will highlight all of them.}
%%% \label{fig:linking}
%%%\end{figure}

\subsubsection{Parallel-Marker Plot for Organ Doses}
While rendering the radiation plans in 3D provides an intuitive understanding of the relationship between the anatomical structure of the patient and the radiation plans, it proved insufficient for understanding the details of how the dose prediction was generated for each organ. Often, the dose distribution will vary significantly in a few organs across the cluster, while others, such as the brainstem and eyes, show little variance. In addition, a small number of matches means that a single outlier can strongly skew the distribution for certain organ predictions. 

As a result, we wanted a way to explore and analyze the dose distribution across the matches used for the prediction, while keeping track of spatially-collocated organs. Because predictions are based on a small number of patients at a time, traditional statistical plots such as box plots or violin plots are not appropriate for this task, as a single outlier would skew them. Likewise, encodings that rely on size to encode distribution density require excessive screen real-estate to be visually discernible, which is infeasible when visualizing a large number (45 organs) of distributions. 

Instead, we introduce a spatially-aware parallel marker encoding to fit our goals (Fig.~\ref{fig:teaser} bottom). The encoding uses a parallel coordinate system, where the x-axis is divided into equal-length bins, each corresponding to one organ of interest in the radiation plan, not including GTVs. To encode spatial organization of anatomical marks, we started by grouping the 45 organs into 6 different categories (Throat, Oral Cavity and Jaw, Salivary Glands, Eyes, Brainstem \& Spinal Cord, and Misc), which were determined after discussion with our radiation oncology collaborators, and we laid out organs within each category contiguously along the x-axis. A vertical line is extended up the center of each bin to provide a visual reference. The order of the axes is fixed and based on the anatomical groups. The y-axis encodes dose, scaled based on the minimum and maximum dose found in the entire dataset. Moving the mouse into a bin highlights the vertical line for that bin, and brings up a tooltip giving the name of the organ, the predicted organ dose, and the actual organ dose for the currently active patient. 

Within each bin, the dose to the specific organ is encoded by one marker per each patient considered for the current prediction. We chose to plot each patient point individually, given the relatively small number of points in each bin. By making makers semi-translucent, regions where several points overlap appear as more opaque, giving a visual indicator of density. The current patient is denoted by a different shape (cross), while matches are shown as dots and colored based on their clusters, maintaining consistency in color and shape with the encodings in the scatterplot. The predicted dose is also denoted by a cross marker, colored in blue. The size of dot markers is based on the computed similarity with the given patient. This encoding serves as a visual metaphor, as larger dots carry more 'weight' in the prediction, and the predicted dose is effectively at the center-of-mass of the dots in each bin. We converged to this composite encoding after experimenting with and discarding parallel coordinate plots, as well as a variety of other axis encodings, markers, and channels.

The different views are linked through color, marker shapes, and brushing and linking. For example, when the user hovers over the encoding of another patient, all other encodings related to the same patient are highlighted in white (Fig.~\ref{fig:teaser}). Additionally, the user can select a patient to bring into focus by clicking on a point in the scatterplot or clicking on the patient ID label above their radiation plan. The data processing and algorithm for our system was implemented in python, using the NumPY library ~\cite{Stefan_Van_Der_Walt_et_al_2011} for doing numerical computations, and Pandas~\cite{Wes_McKinney_2010} for data-processing. The front-end visualization was implemented as a web-based tool using HTML, CSS, and Javascript, with the three.js~\cite{cabello2010three} and d3.js~\cite{bostock2011d3} libraries.

\begin{figure}[t]
 \centering 
 \includegraphics[width=\columnwidth, height=95mm]{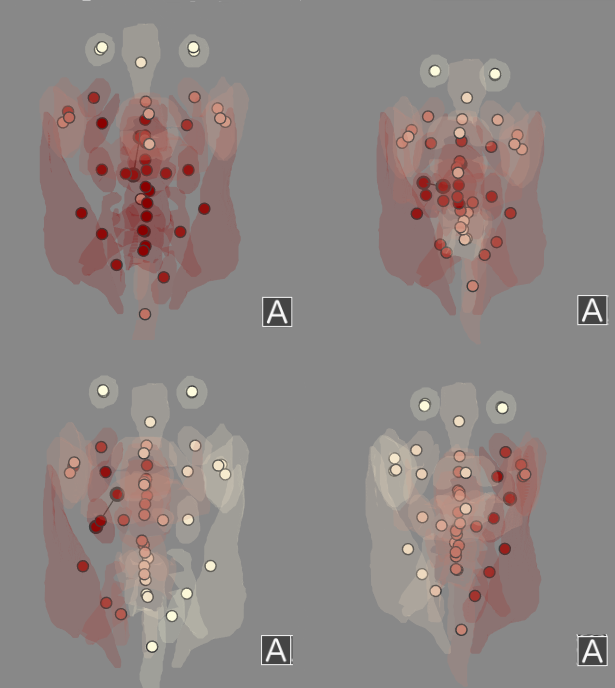}
 \caption{Example radiation plans for the 4 different patterns identified in the data. Top left: a plan with a higher dose to the lower-anterior throat. Top right: a plan with a 'standard' dose distribution, where radiation is lower in the throat and distributed to both the left and right sides of the head. Bottom right: a plan with dosing primarily to the right side of the head. Bottom left: a plan with dosing primarily to the left side of the head.}
 \label{fig:rtpatterns}
\end{figure}

\begin{figure*}[ht]
 \centering 
 \includegraphics[width=\textwidth, height = 105mm]{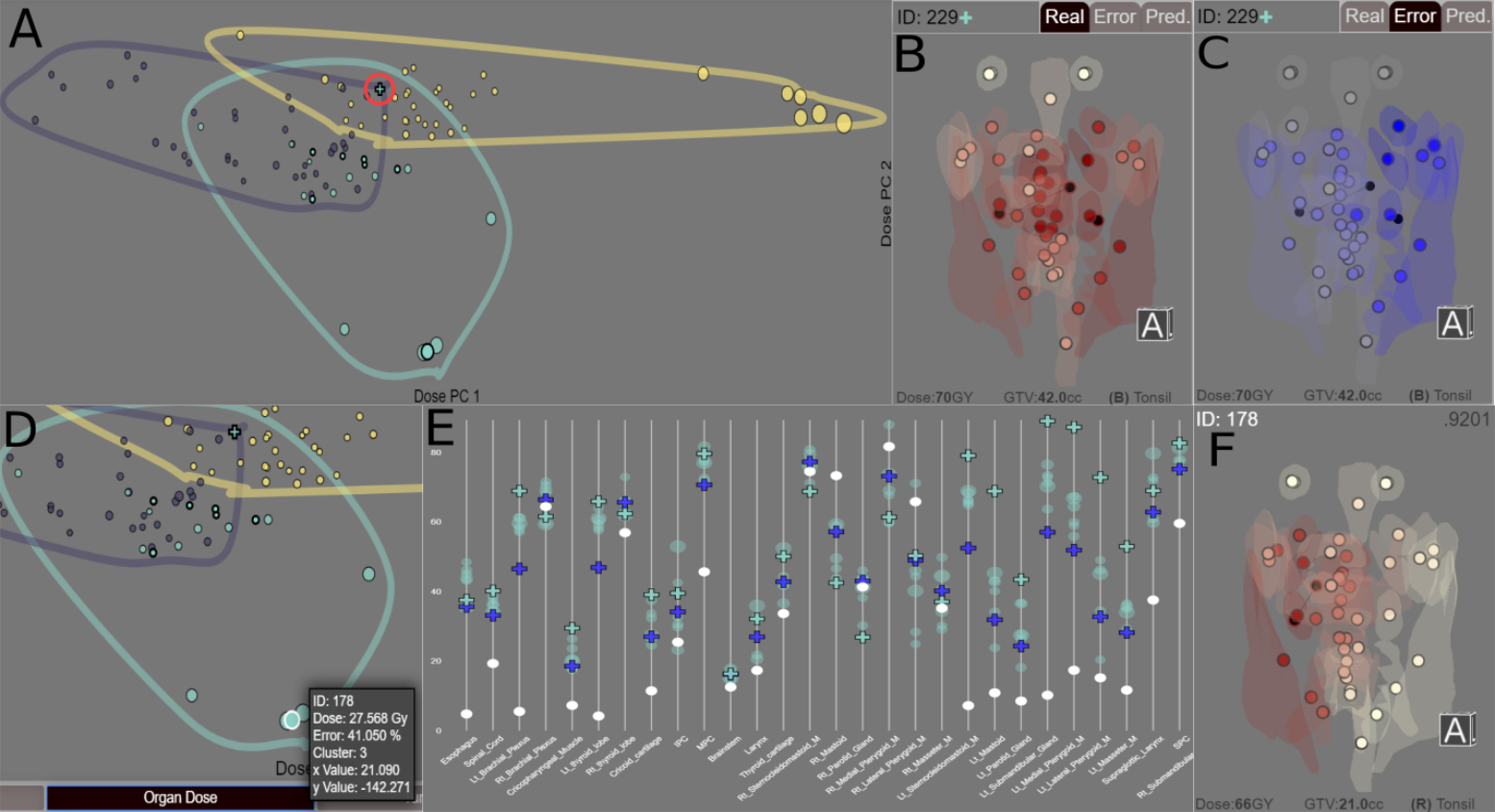}
 \caption{Snapshots of key moments during the qualitative evaluation. (A) Picture of the dose-PCA scatterplot on the reduced cohort using the clustering provided by GC. Clusters visibly divide the feature space despite being done without dose information. (B) RT plan for the patient being inspected (shown in (A) as the cross cyan marker, circled here in red). (C) RT prediction error for the patient. Error rates are highest on the left side of the head. (D) Close up of the dose-distribution. One of the matches (highlighted) is significantly further from the other matches. (E) Parallel-marker dose plot of the patient and its matches. Doses from the suspicious match (highlighted) are significantly lower for several adjacent areas. (F) Radiation plan of the suspiciously-matched patient, who, despite a similar tumor location, received almost no radiation to the left side of their head.}
 \label{fig:evaluation}
\end{figure*}

\section{Evaluation and Results}
\label{sec:evaluation}
Because of the visual computing nature of this project, we use a hybrid quantitative and qualitative evaluation methodology shaped along two case studies. We first present a case study of how visual analysis was used in conjunction with our similarity measure to help develop and improve the prediction algorithm. Along with this discussion, we present quantitative data about the prediction performance. In the second case study, we present a qualitative evaluation done with four senior domain experts in data mining, biostatistics, cancer medicine, and medical imaging.

\subsection{Case Study: Algorithm Development}
One of the topics of interest to our collaborators was understanding the importance of structural similarity in predicting radiation plans. However, traditional prediction methods are complicated by the fact that radiation plans can vary widely based on subjective planning factors that can be patient-case, clinician, or institution specific. In this first analysis, we discuss the development and performance of our prediction algorithm in conjunction with this goal, demonstrate how insight from the visual computing tool was leveraged to help improve the prediction algorithm, and how visualization can be used to convey the results to clinicians, to allow for better expert feedback in the algorithm design process.

We begin by first describing our measure for quantitatively assessing the success of the prediction algorithm. Given that for each patient in the cohort we have access to the actual RT plan for that patient, the accuracy of prediction across the cohort can be computed via leave-one-out validation, as follows: 1) for each patient in the cohort, use the tumor-to-organ distances and organ volumes to determine the most similar patients in the cohort via the T-SSIM similarity measure; 2) use the set of similar patients' RT plans to predict the dose distribution per organ (i.e. the RT plan) of the current patient; 3) compute and report the prediction error as the difference between the predicted RT plan and the actual RT plan for that patient; 4) report the mean error across the cohort. In assessing error, we chose to compute the total absolute error for each patient. We decided on this measure over root mean squared error (RMSE), because RMSE is typically used to more strongly punish outliers. %Because we are interested in typical patterns for the patient, we are less interested in the effect that outliers have on the prediction. 

Using the similarity measure and prediction algorithm without dividing the cohort into clusters, we initially found a mean prediction error of 16.68\%, or 6.15 Grays (Gy), with a standard deviation of 9.31\%. We compared this method to the naive method, where the predicted dose distribution is simply the average of the entire cohort. Using this naive method, we get a mean error of 20.62\%, or 7.48 Gy with a standard deviation of 14.0\%, which was suspiciously close to the performance of our initial prediction. 

To better understand these results, the data and outliers were inspected using the visual steering tool. For each outlier prediction in the dataset, we inspected the k nearest neighbors selected for the prediction in the RT panel adjacent to the outlier patient. Where visual inspection did not pick up on subtle cues, the dose distribution plot was particularly useful in helping identify suspicious neighbor matches. Using the RT views, RT outliers were found to belong in three distinct pattern classes. Patients in these classes had larger errors, suggesting that they had peculiarities in their dose distribution that were not being captured by our similarity measure alone. RT plans for the patients in the 3 classes were analyzed and discussed with our radiation oncology collaborators and contrasted with patients with good predictions. In this manner, we identified four distinct patterns in how the RT plans were distributed (Fig.~\ref{fig:rtpatterns}). This finding was subsequently confirmed in the scatterplot panel. The first, largest group was the 'standard' plan, recognized by our collaborators as most common for the cohort. Another group comprised a subset of the patients that received additional radiation to their lower throat, near the larynx. While surprised by this finding, our collaborators found this second RT plan type consistent with results reported by Amdur et al.~\cite{Robert_J._Amdur_et_al_2004}. Amdur et al. discussed the choice of delivering additional irradiation to the larynx in patients and compared it to other methods of irradiation that largely avoid irradiating the larynx at all, leading to two potentially highly different dose distributions based on subjective choices made by the physician. The remaining two plan types were groups that appeared to have received highly unilateral radiation to only a specific side of their head, with the two groups corresponding to the two sides of the head. The radiation oncologists were enthusiastic and surprised by the power of the measure in making these findings possible. It was determined after discussion with our collaborators that the differences between the four plan types were likely due to radiation planning methods related to several other factors than tumor location, including the health of the patient, the tumor staging, and whether a biopsy had previously been done on the primary or secondary tumor.

Given this insight, we investigated introducing four clusters into the prediction, based on the different radiation plan archetypes found. This time, by only considering similar patients within the same cluster, our prediction error dropped to 12.3\%, or 4.71 Gy, with a standard deviation of 4.43\%. When normalized by prescribed dose, the total prediction error is 6.87\% across the four clusters and for the 45 organs considered.  Beyond the ability of the measure to identify the four RT classes, this prediction power was considered remarkable by our medical collaborators.

\subsection{Case Study: Toxicity and Clustering Outlier}
Because our project aims to support expert researchers in a specialized domain, we performed a remote qualitative evaluation with four senior domain experts, who are co-authors on this paper (GC, DV, BE, GM). The experts have backgrounds in data mining, biostatistics, radiation oncology, and medical imaging, respectively. All participants were familiar with the visual computing application throughout its development stages. Because of the experts' participation in the design process, the lack of an alternative existing system to solve the same problem, and in further accordance with the ACD paradigm, the evaluation was focused on the functionality of the application with respect to the target problem. Participants were given a briefing on the different components and basic functionality of the visual interface and were encouraged to ask questions to guide the exploration of the data and results. The first author navigated the application with direct guidance from the participants, who were shown the same screen and were able to communicate with each other. 

The main goal was to investigate whether our similarity measure can predict whether a patient will develop a particular toxicity (side-effect) after RT treatment, such as requiring the insertion of a feeding tube (FT). There are no current algorithms that can accomplish this type of prediction. The starting point of this investigation was a subset of 92 patients in the cohort for whom toxicity data was readily available. Collaborator GC had generated a clustering of this subset using our similarity measure, with the aim of correlating the tumor-locations and RT plans with the toxicity data. The clustering had yielded three clusters, one of which was statistically correlated with the feeding tube toxicity. 

The investigation (Fig.~\ref{fig:evaluation}) started with the group examining the resulting clusters. Clustering had been done on the patient similarity scores provided by our similarity measure, and no expert (including GC) had seen the labeled results before in the context of the patient spatial information. The analysis started with the scatterplot visualizing the clusters, followed by targeted questions about the three PCA tabs. In the organ-dose plot, a collaborator noted that the clustering visibly divided the patients into separated groups. This was exciting to the group, given that the clustering had been done over the spatial similarity only, independent of dose. One of the visual computing researchers pointed out the cluster that was statistically correlated with the feeding tube outcome (navy cluster in Fig.~\ref{fig:evaluation}.A). 

Upon further inspection, the group noted that some of the matches within a different cluster (cyan cluster in Fig.~\ref{fig:evaluation}.A) were far apart in the organ-dose plot, while being close in tumor-organ distance plot. The group asked why that was, and proceeded to examine the RT views of that patient (Fig.~\ref{fig:evaluation}.B), followed by the patient's predicted RT plan. Upon noticing spatially-localized higher prediction errors (Fig.~\ref{fig:evaluation}.C), the group proceeded to examine the RT views of the nearest neighbors who had been used to compute the prediction. By linking the view of each neighbor with the corresponding highlighted mark in the organ-dose scatterplot, the group was able to determine a suspicious match: while the tumor location in the neighbor was very similar to the one in the patient under consideration, the two patients were far apart in the dose-distribution plot (Fig.~\ref{fig:evaluation}.D). A detailed investigation of the two patients and their matching followed, this time using the parallel-marker plot (Fig.~\ref{fig:evaluation}.E). One of the experts noted a localized difference in a contiguous subset of organs in the marker plot (last quartile of x-axis), and as the group circled back to the RT view of the match, they noticed that the neighbor's RT plan featured a low dose to half of the head (Fig.~\ref{fig:evaluation}.F). The expert in radiation oncology explained that the way the radiation plan was done could have been affected by a number of factors, e.g., whether a biopsy had been performed on the patient's lymph node. This led to a group discussion of the earlier case study and the usefulness of including a fourth cluster in the analysis, potential ways to incorporate more patients, and future plans to predict other toxicity outcomes based on the RT prediction.

An interesting result of this evaluation was the ability of the different domain experts to guide parts of the visualization and ask questions to each other. The collaborator with a background in data mining understood principal component analysis, and was able to explain the plot tabs to another expert. Instead of stopping the investigation with a convenient p-value finding, the group continued to examine the clustering that had generated that outcome, and were able to spot outliers and suggest improvements to the clustering. The medical imaging expert caught on the spatial dose pattern and explained it to the other specialists. When analyzing why two patients were being matched despite having notably different dose profiles within the clustering, the expert in radiation oncology provided the rest of the group with a clinical rationale for that fact. The statistician picked up on that interpretation, and suggested additional data collection. The group was able to efficiently use the whole system in order to make an important observation. Overall, we believe that this evaluation highlights a potential for visual computing methods such as these to support interdisciplinary collaboration more effectively.
%, by combining backgrounds in traditionally different fields so that experts in different disciplines can cooperatively analyze the data.

\section{Discussion and Conclusion}
This work introduces a hybrid, topology and image-fidelity, approach to creating an RT spatial similarity measure. Our results show that the resulting measure can successfully retrieve patients with similar tumor location. The similarity measure was then successfully used to make a valid prediction of RT dose distributions in a new head and neck cancer patient. The development of this measure and prediction algorithm was made possible through a visual steering approach, where a visual interface coupled with the spatial algorithms enabled us to identify and analyze situations where early algorithm versions failed. The same approach enabled us to identify four specific RT patterns in the data, and, in conjunction with the spatial similarity measure, to improve prediction. When evaluated on a dataset of 165 patients, the prediction had low mean error: 4.71 Gy, compared to doses per organ as high as 70-90 Gy. We also observed low 4.43\% standard deviation in the computed error, suggesting high certainty in our prediction. This type of certainty is particularly important when dealing with life-affecting patient outcomes. In conjunction with clustering, the spatial measure enabled detecting correlation between patient groups and a specific toxicity, paving the way towards precision medicine that leverages spatial information in patient data repositories.

Another result of this integrated approach is the ability to visually assess outliers and problems in the data. Since our data relies on segmentation of complex CECT images, problems in the data are to be expected. The high-dimensional nature of this data, combined with a relatively small dataset, makes outlier detection using traditional methods difficult. Additionally, automatic outlier detection methods are insufficient, since the presence of different clusters in the radiation plans means that new data could appear to be outliers, when in fact they are valid, but uncommon, RT plans, or that bad data can insidiously look 'normal'. However, by visualizing outliers, we were able to consult with experts in order to determine if the resulting anatomies and radiation plans are plausibly valid, or can be removed. For example, two patients in the cohort had several organs, including their eyes, positioned near the base of their throat. While these configurations are physiologically impossible, they were not detected in standard outlier detection, and even showed high similarity scores with each other.

Our qualitative evaluation also shows that an approach grounded in the ACD paradigm and visual scaffolding principles can lead to a satisfactory outcome for a difficult scientific problem. Using this approach, collaborators with a variety of complementary expertise were able to work together in order to gain insight into the relationship between spatial information and RT plans. A coordinated-views paradigm allowed us to leverage visual representations familiar to some of the experts, in order to expose those experts to novel or unfamiliar encodings. For example, oncologists were able to make connections between RT volume renderings and the cluster and parallel-marker encodings. In the same vein, we note that our parallel-marker plot builds on familiar statistical plots while accommodating fewer samples and spatial contiguity. Because these visual encodings were developed through participatory design, we do not explicitly report feedback, which was enthusiastic, from our collaborators. 

While our approach and spatial algorithms are generalizable to other problems in medicine and elsewhere, we note that there are limitations as well. First, details of an RT plan can change based on factors specific to the clinician and institution. For example, we have seen in our data that there are many cases where two patients are similar in terms of tumor location, but one patient has highly-unilateral dosing. When generalizing a prediction method, we have to consider that other clusters could arise due to differences in the data, as well as technological and methodological differences between institutions. As a result, being able to inspect the data and leverage clinical knowledge is an essential function that can be accomplished through the use of visual computing.

Furthermore, while our current measure can encapsulate volumetric and spatial information, microscopic as well as higher-level information on the organ structure, such as shape and orientation, could be relevant and could also be included. Additionally, computing the similarity scores requires $\mid Cohort \mid ^{2}$ computations, where $\mid Cohort \mid$ is the size of the cohort, and so it is done offline, while more sophisticated clustering methods are run off-site. This means that currently, online analysis can take place only once the results are generated.  On 5 trials using a machine with 8GB DDR4 RAM and Intel i5-7200U 2.5GHz processor, the offline calculation took under 10 minutes (100.5s for processing and 476.5s for prediction, on average).  This amount is negligible compared to the week-long IMRT planning process, which also requires medical professional input during the planning. In addition, our parallel-marker plot, which works well for 45 organs, has limited scalability to thousands of measurements. Finally, while our approach does not rely on learned parameters, we need to specify two meta-parameters: window size for computing organ adjacency, and an optimal number of matches to use in the prediction,  which may affect generalization. In our study, we found the optimal parameters via simple optimization~\cite{swann1969survey}. 

In conclusion, we presented a visual computing approach to support the development of a predictive algorithm to estimate radiotherapy plans in head and neck cancer patients. We introduced a novel, hybrid way of measuring anatomical similarity based on topology and measures of image fidelity. This similarity measure was then used in the emerging field of precision oncology, to retrieve patients in a cohort who are likely to have similar radiation plans and outcomes. By tightly coupling a visual analysis interface and a novel encoding with our algorithms, we derived valuable insight into the role that spatial information plays in radiation therapy planning, and were able to drive the development of the predictive algorithm. This visual steering approach is supported by coordinated views of spatial and nonspatial, statistical data. These views allowed domain experts in radiation oncology, statistics, data management and medical imaging to explore the data from different perspectives. Ultimately, the visual computing methodology presented in this paper enables calculations and insights into medical data that were otherwise not possible. 

\section*{Acknowledgements}
This work was supported by the National Institutes of Health [NCI-R01-CA214825, NCI-R01CA225190] and the National Science Foundation [CNS-1625941, CNS-1828265]. We thank all members of the Electronic Visualization Laboratory, members of the MD Anderson Head and Neck Cancer Quantitative Imaging Collaborative Group, and our collaborators at the University of Iowa and University of Minnesota. 
\bibliographystyle{abbrv-doi}
\bibliography{bibliography}

\end{document}